\documentclass{rspublic}

\usepackage{graphics}           
\usepackage{bm}                 
\usepackage{amsmath}            
\usepackage{amsfonts}
\usepackage{amssymb}
\usepackage{latexsym}           

\usepackage[round,colon,authoryear]{natbib}
\bibpunct[, ]{(}{)}{;}{a}{}{,}
\newcommand{\ket}[1]{\mbox{$|#1\rangle$}}

\newcommand{\identity}{\leavevmode\hbox{\small1\kern-3.2pt\normalsize1}}

\begin{document}

\title
      [Quantum random walk algorithms]
      {A random walk approach to quantum algorithms}

\author
       [V.~Kendon]
       {Vivien M.~Kendon}

\affiliation{School of Physics and Astronomy, University of Leeds, Leeds LS2 9JT, UK}

\label{firstpage}

\maketitle

\begin{abstract}{quantum information, quantum computation, quantum algorithms}
The development of quantum algorithms 
based on quantum versions of random walks is placed in the
context of the emerging field of quantum computing.
Constructing a suitable quantum version of a random walk is not trivial:
pure quantum dynamics is deterministic, so randomness only
enters during the measurement phase, i.e., when converting the
quantum information into classical information.
The outcome of a quantum random walk is very different from the
corresponding classical random walk, due to interference between
the different possible paths.
The upshot is that quantum walkers find themselves further
from their starting point on average than a classical walker,
and this forms the basis of a quantum speed up 
that can be exploited to solve problems faster.
Surprisingly, the effect of making the walk slightly
less than perfectly quantum can optimize the properties of
the quantum walk for algorithmic applications.
Looking to the future, with even a small quantum computer available,
development of quantum walk algorithms might proceed more
rapidly than it has, especially for solving real problems.
\end{abstract}

\section{Introduction}
\label{sec:intro}

The idea that quantum systems might be able to process information 
fundamentally more efficiently than our everyday classical computers
arose over twenty years ago from visionary scientists such as
\cite{feynman82a} and David \cite{deutsch85a}.  They both 
perceived that a superposition of multiple quantum trajectories looks like
a classical parallel computer, which calculates the result
of many different input values in the time it takes for one processor
to do one input value.  Except the quantum system doesn't need a stack
of processors, the parallel processing comes `for free' with 
quantum dynamics, providing a vast economy of scale over classical
computation.  For the next ten years this remained a neat idea
with no hope of practical application, since the fragile nature of
quantum superpositions could not be perfectly controlled to yield
a functional quantum computer.  Then came two milestone results in
quantum computation: error correction 
\citep{knill96a,aharonov96a,steane96a} to protect the fragile
quantum systems long enough to run a computation, and Shor's algorithm
\citep{shor95a}
for factoring large numbers, which threatens the security of cryptographic
systems.  This gave quantum computing both practicality and purpose: the
growth in research over the past ten years has been phenomenal on both
theoretical and experimental aspects of the challenge to construct a
working quantum computer.

A quantum computer requires both hardware and software.  Both 
are proving to be fascinating areas of research, with many beautiful
results appearing at each step of the process.
In terms of size, the hardware is not very far advanced:
we cannot even manipulate as many as ten qubits (quantum bits) long enough
to do a simple calculation a child could do in their head.  But this
belittles the exquisite control over single particles that has been developed
in a diverse set of fields from photons to trapped atoms and ions
to quantum dots to electrons floating on liquid helium to nuclear spins
to SQUIDS (superconducting quantum interference devices), see
\cite{spiller05a} for a recent review.
Progress has
been steady and impressive: nothing has yet been discovered in all
these experiments that says we cannot expand them as far as is
necessary to make a useful quantum computer.

The software side is also at a relatively early stage.  Shor's factoring
algorithm is the first in a family of quantum algorithms based on Fourier
transforms, which exploit the fact that a quantum computer can calculate a
Fourier transform efficiently for many different inputs in superposition
then extract a common periodicity from the result.  In technical terms,
the task is to identify a hidden subgroup in the group structure of
the problem \citep[provides a recent review]{lomont04a}.
This works well for Abelian groups, but extending
the method to a non-Abelian groups, where some of the notorious hard 
problems, such as graph isomorphism, lie, is proving tricky.
The parallelism of quantum systems may come `for free', but extracting
the answer does not.  When a quantum superposition is measured, it only
gives out a single randomly selected result from the many it is holding.
One therefore has to be very ingenious about how to arrange the
superposition prior to the measurement in order to maximize
the chance of obtaining the required information.

Clearly we'd like to find many more methods for
programming a quantum computer.  One obvious place to look 
is where classical algorithms are having the most success, to
see if a quantum version could be even faster.  Randomized algorithms
are one such arena, they provide the best known methods for approximating
the permanent of a matrix \citep{jerrum01a}, finding satisfying
assignments to Boolean expressions ($k$SAT with $k>2$) \citep{schoning99a},
estimating the volume of a convex body \citep{dyer91a},
and graph connectivity \citep{motwani95}.
Classical random walks also underpin many standard methods in computational
physics, such as Monte Carlo simulations.
There are two approaches to quantum versions of random walks: \cite{farhi98a}
investigated quantum walks using a discrete space (lattice or graph) with
a continuous-time quantum dynamics,
and \cite{aharonov00a}
studied quantum walks with both space and time discretized.
Within a few years, the first proper algorithms using quantum walks 
appeared from \cite{childs02a} and \cite{shenvi02a}, and more have since
followed: for a short survey see \cite{ambainis04a}.

In this paper we will review how the simplest quantum walk behaves
in comparison with a classical random walk, then briefly describe
the two above-mentioned quantum walk algorithms.
We will then turn to more basic questions about how
quantum walks work, including the effects of making them slightly
imperfect.  Finally, we will consider the possible future 
of quantum walk algorithms in the context of the prospects
for practical quantum computing.

\section{A simple quantum walk}
\label{sec:qwalk}

Lets start with a random walk on a line
-- a drunkard's walk -- where the choice of whether to step to the
right or the left is made randomly by the toss of a coin.
A step-by-step set of instructions for a classical random walk is given
in table \ref{tab:recipes} in the left-hand column.  As is well
known, after taking $t$ random steps, the average distance from the
starting point is $\sqrt{t}$ steps.  
\begin{table}
\caption{Recipes for a classical random walk (left) and a quantum random
	walk (right) on a line.
	The symbol $|x,c\rangle$ denotes a quantum walker at position $x$
	with a coin in state $c$.  The quantum operations $\mathbf{H}$ and
	$\mathbf{S}$ are defined by their effect on $|x,c\rangle$ as given below.}
\begin{center}
\begin{tabular}{ll}
\textbf{Classical random walk} & \textbf{Quantum walk} \\
1.~start at the origin: $x=0$ & 1.~start at the origin: $x=0$ \\
2.~toss a coin & 2.~toss a qubit (quantum coin)\\
\null\hspace{2em}result is \textsc{head}  or \textsc{tail}
 & \null\hspace{2em}$\mathbf{H}|x,0\rangle\longrightarrow (|x,0\rangle + |x,1\rangle)/\sqrt{2}$\\
 & \null\hspace{2em}$\mathbf{H}|x,1\rangle\longrightarrow (|x,0\rangle - |x,1\rangle)/\sqrt{2}$\\
3.~move one unit left \textbf{or} right &
    3.~move one unit left \textbf{and} right\\
\null\hspace{1em}according to coin state:
 & \null\hspace{1em}according to qubit state\\
\null\hspace{2em}\textsc{tail}: $x\longrightarrow x-1$
 & \null\hspace{2em}$\mathbf{S}|x,0\rangle\longrightarrow |x-1,0\rangle$\\
\null\hspace{2em}\textsc{head} : $x\longrightarrow x+1$
 & \null\hspace{2em}$\mathbf{S}|x,1\rangle\longrightarrow |x+1,1\rangle$ \\
4.~repeat steps 2. and 3. $t$ times  & 4.~repeat steps 2. and 3. $t$ times \\
5.~measure position $-t \le x \le t$ & 5.~measure position $-t \le x \le t$ \\
6.~repeat steps 1. to 5. many times & 6.~repeat steps 1. to 5. many times \\
$\longrightarrow$ prob. dist. $P(x,t)$, binomial &
    $\longrightarrow$ prob. dist. $P(x,t)$ has \\
standard deviation $\langle x^2\rangle^{1/2} = \sqrt{t}$ &
    standard deviation $\langle x^2\rangle^{1/2} \propto t$ \\
\end{tabular}
\label{tab:recipes}
\end{center}
\end{table}
Na\"{i}vely, for a quantum version of this based on the Feynman/Deutsch
view of quantum computation as parallel processing,
one would expect to follow all the possible random walks in superposition,
instead of just one particular sequence of steps.  
At each step, then, move left and right in equal amounts, repeating this
until the quantum walker is smeared out along the line in the same
distribution as one finds after many trials with a classical random walk.

\begin{figure}
  \begin{center}
    \resizebox{0.7\textwidth}{!}{\includegraphics{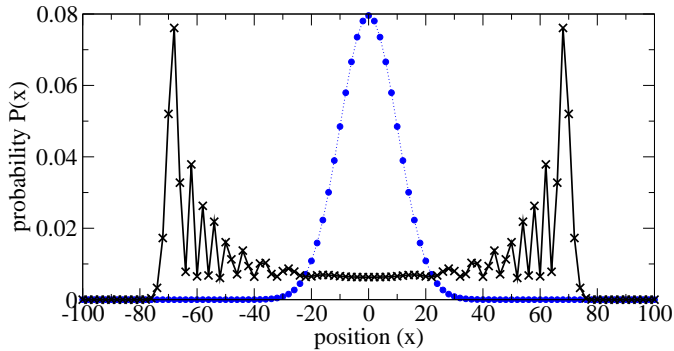}}
  \end{center}
  \caption{Comparison of quantum (crosses) and classical (circles) walks
	on a line, after 100 steps. Only even points are shown since odd
	points are unoccupied at even time steps.}
  \label{fig:qvscwalk}
\end{figure}

However, this isn't possible within the constraints of
quantum dynamics, as shown by \cite{meyer96a}. Pure quantum
dynamics must be unitary, which means being completely deterministic
and reversible.  The na\"{i}ve superposition of all possible random
walks is not reversible, 
you can't tell which way you arrived at a particular location,
so you don't know which way to step to go back.
The solution
is to make the coin quantum too, and give it a unitary twist instead of
a random toss.  The quantum coin keeps track of which way you arrived at
your location, allowing you to retrace your steps.
This recipe for a quantum walk was first investigated with algorithmic
applications in mind by \cite{aharonov00a} and \cite{ambainis01a}.
Table~\ref{tab:recipes} compares the instructions for performing classical and
quantum random walks on a line.
The shape of the distributions obtained is shown in figure \ref{fig:qvscwalk}.
The quantum walk looks nothing like the binomial distribution of the
classical random walk.  But we have gained something significant: the
quantum walk spreads faster, at a rate proportional to the number of steps,
instead of the square root of the number of steps as in the classical case,
a quadratic speed up.

\section{Quantum walk algorithms}
\label{sec:qwalgos}

Of course, quadratically faster spreading on a line isn't an algorithm yet,
but it is a good start, and soon \cite{shenvi02a}
showed that a quantum walk could search an unsorted database
with a quadratic speed up.  The first quantum algorithm for this
problem is due to \cite{grover96a}, using a different method to 
obtain the same quadratic speed up.
A classical search of an unsorted database (e.g., starting with a phone
number and searching a telephone directory to find the corresponding name),
potentially has to check all $N$ entries in the database, and on average
has to check at least half.  A quantum search only needs to make
$\sqrt{N}$ queries, though the queries ask for many answers in superposition.
The quantum walk search algorithm sort of works backwards, starting in
a uniform superposition over the whole database, and converging on the
answer as the quantum walk proceeds.
As already noted, quantum walks are reversible: a quantum walk running
backwards is also a quantum walk.

\begin{figure}
    \begin{center}
    \resizebox{0.5\textwidth}{!}{\includegraphics{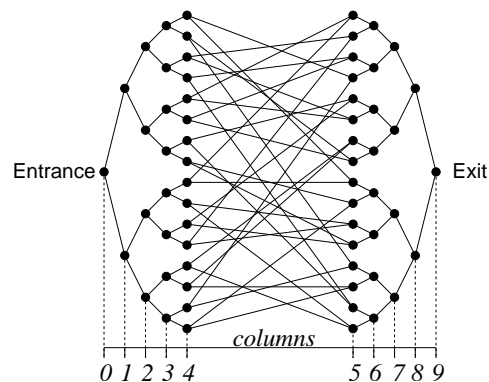}}
    \end{center}
    \caption{(Small) example of the network in \cite{childs02a}. The gap
	between columns 4 and 5 is for clarity only, the length of the
	paths is the same between all nodes.}
    \label{fig:tree4}
\end{figure}

\cite{childs02a} proved that a quantum walk
could perform exponentially faster than any classical algorithm when 
finding a route across a particular sort of network,
see figure \ref{fig:tree4}.  This is a rather artificial problem,
but proves in principle that quantum walks are a powerful tool.
The task is to find your way from the entrance node to the exit node,
treating the rest of the network like a maze where you cannot see the other
nodes from where you stand, only the choice of paths.  It is easy to tell
which way is forward until you reach the random joins in the centre.
After this, any classical attempt to pick out the correct path to the exit 
get lost in the central region and takes exponentially long, on average,
to find the way out.  A quantum walk, on the other hand,
travels through all the paths in superposition, and the quantum interference
between different paths
allows the quantum walker to figure out which way is forward right up to
the exit, which it finds in time proportional to the width of the network.
\cite{childs02a} use a continuous time quantum walk for this algorithm.
The recipes for continuous time walks are similar to the discrete time
walks in table \ref{tab:recipes}. Continuous time walks do not
use a coin: the coin toss and conditional shift
(steps 2.~and 3.) are replaced by a hopping rate $\gamma$ per unit time
for the probability of moving to a neighbouring location,
and step 4.~by continuous evolution for a time $t$,
see \cite{farhi98a}.

\section{Decoherence in quantum walks}
\label{sec:decoher}

Our quantum walks do not involve any random choices, and
they have very different outcomes to classical random walks, so
why are they a good quantum analogue of a classical random walk?
The similarity of the recipes in table \ref{tab:recipes} should convince
you that the choice is at least a plausible one.  One might also 
reasonably require that the quantum walk reduces to the classical walk
`in the classical limit'. By this we usually mean the limit
obtained by making the system bigger and bigger,
until Planck's constant $h=6.6\times 10^{-34}$Js
is very small compared to the energy and time scales
in the system.
We can't apply this directly to our quantum walks, but another way to
transform to a classical dynamics is by applying decoherence to
destroy the quantum correlations that allow the
quantum superpositions to exist.

Decoherence is essentially an interaction with the environment, in which
the correlations within the system are transformed into correlations with
the environment that we can no longer access or control.  We can
mimic this by measuring our quantum system: this correlates it with our
measuring apparatus, but since the measuring apparatus is big enough to
be classical in the sense of $h\rightarrow 0$,
it forces our quantum walker
to a classical state too, removing any quantum superpositions.
If we measure the walker after each step, we will find it in just one location
on the line, and because the quantum coin toss gives equal probability
of moving left and right, we will find the walker one step to the left
or right of where it was when we measured the previous step, with equal
probability.  This is exactly the recipe for a classical random walk,
so our quantum walk recipe passes the test.
In fact, this is the best definition we have of a
quantum walk: a quantum dynamics that reduces to a classical random
walk when completely decohered.  There are many different
examples of classical random walks, and no convenient quantum definition
is known that covers everything that reduces to them.
One may even start from classical random walks and quantize them 
\citep[see, for example,][]{szegedy04a}.

In this sense, then, decoherence is absolutely basic to the definition
of a quantum walk.  But in a practical sense, decoherence is irrelevant
for quantum walk algorithms. 
When constructing a physical quantum computer,
we have to worry a great deal about decoherence, and build in enough
error correction to allow the quantum computation to proceed without
any errors.  If we run our quantum walk algorithm on a quantum computer,
just as with our classical computers, we should expect a properly
functioning quantum computer to carry out the algorithm without any
mistakes due to the action of the environment on the physical computer.

This is not the end of the story on decoherence in quantum walks.
When asked whether decoherence would spoil a quantum walk,
I thought: silly question, but, somehow that spiky quantum
distribution has to turn into the smooth classical one: I wonder how 
it does it?  The answer is in figure \ref{fig:qwdec}:
\begin{figure}
  \begin{center}
    \resizebox{0.7\textwidth}{!}{\includegraphics{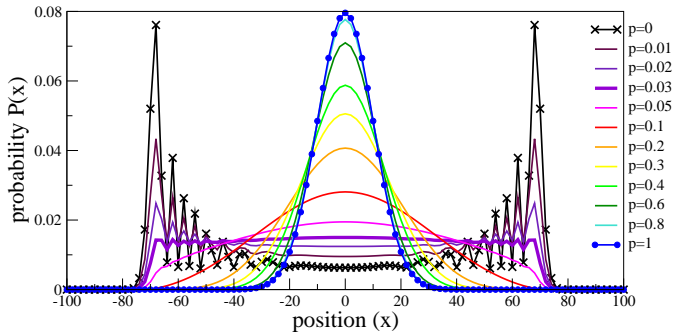}}
  \end{center}
  \caption{A quantum walk on a line of 100 steps is progressively
	decohered by random measurements with probability $p$ per
	time step as given in the key.  for $p=0.03$, an approximately
	`top hat' distribution is obtained \citep[from][]{kendon02c}.}
  \label{fig:qwdec}
\end{figure}
down come the spikes and up goes the central peak as the decoherence rate
is turned up.  And in between it passes through a good
approximation of a `top hat' distribution,
which, for computational physicists who use random walks to sample
distributions, is a very desirable feature, since it provides uniform
sampling over a specific range.

Inspired by this result, \cite{kendon02c}
also checked what happens to quantum walks
on cycles.  These are what you get if you take a line segment of length
$N$ and join the ends to make a loop.  The quantum walk goes round and
round in both directions and, unlike a classical random walk, which 
settles down to within epsilon of a uniform distribution after a
certain length of time
-- the mixing time -- a quantum walk continues to oscillate indefinitely.
Mixing is needed for uniform random sampling.  The shorter
the mixing time the more frequently you can sample, and the quicker you
obtain a good estimate of the answer to your problem.  If quantum
walks don't mix at all, this seems to be useless for algorithms.
Fortunately, there is a simple remedy: we define an average distribution
\begin{equation}
\overline{P(x,T)}=\frac{1}{T}\sum_{t=1}^{T} P(x,t).
\label{eq:meanP}
\end{equation}
Operationally, this just means randomly choosing a value of $t$ between $1$
and $T$, then measuring the position of the quantum walker after $t$
steps instead of after $T$ steps.  This time-averaged distribution
does settle down after an initial mixing time.  It does not necessarily
settle to the uniform distribution, though \citep{aharonov00a},
unlike the classical random walk which always mixes to a uniform distribution.
This is thus another striking difference between quantum and classical
random walks.  However, if one uses odd-sized cycles, then the quantum
walk does mix to uniform, and \cite{aharonov00a} proved that they mix
almost quadratically faster than classical random walks (in time proportional
to $N\log N$ steps or better, compared to $N^2$ steps).
Applying decoherence to a quantum walk on a cycle causes it to mix even
faster than a pure quantum walk,
and to always mix to a uniform distribution \citep{kendon02c}.
So again, like the top hat distribution, decoherence seems to enhance
the useful quantum features.
Recent work by \cite{richter06a} suggests that the enhancement is linked
to an amplification technique first applied to quantum walks by
\cite{aharonov00a}.

However, quantum walks on other structures, such as the hypercube,
do not necessarily mix faster than classical random walks, \citep{moore01a}.
Such effects may be dependent on the structure having a favourable symmetry
\cite{krovi06a,keating06a}.
The real lesson to take from this is that we do not have to
be restricted to pure quantum dynamics when designing algorithms,
we have far more tools at our disposal to tune and optimize our
quantum walk to suit the problem at hand.  
Indeed, the idea that measurements form an integral part of
the quantum computational process has been suggested in several
contexts \citep[for an overview see][]{jozsa05a},
in particular the cluster state model by \cite{raussendorf01a},
And in practical situations, a constant factor of (say) 100 makes a big
difference to what can be achieved with a computer of fixed size, so 
algorithmically insignificant effects may still be valuable once usable
quantum computers are available.

\section{What makes a quantum walk `quantum'?}
\label{sec:complement}

How does the quantum walk `go faster' than a classical random walk?
Those readers familiar with wave dynamics (in whatever context) will
have recognized the interference that is occurring between
different paths taken by the quantum walker.  Here it is in three steps
of the quantum walk on a line.
Our notation is the same as in
table~\ref{tab:recipes}, with $\ket{x,c}$ denoting a quantum walker
at location $x$ with a quantum coin in state $c$.  As defined in
table~\ref{tab:recipes}, $\mathbf{H}$ is the quantum coin toss operator,
and $\mathbf{S}$ is the quantum step operator:
\begin{eqnarray}
\mbox{t=0\hspace{1ex}} & 	&\mbox{\null\hspace{2.2em}} \ket{0,0}\nonumber\\
\mbox{t=1\hspace{1ex}} & \mbox{apply $\mathbf{H}$} &
	\longrightarrow(\ket{0,0} + \ket{0,1})/\sqrt{2}\nonumber\\
 & \mbox{apply $\mathbf{S}$} 
	& \longrightarrow(\ket{{-1},0} + \ket{1,1})/\sqrt{2}\nonumber\\
\mbox{t=2\hspace{1ex}}
	& \mbox{apply $\mathbf{H}$}
	& \longrightarrow(\ket{{-1},0} + \ket{{-1},1}
			       + \ket{1,0} - \ket{1,1})/2\nonumber\\
 & \mbox{apply $\mathbf{S}$}
	& \longrightarrow(\ket{{-2},0} + \ket{0,1}
			       + \ket{0,0} - \ket{2,1})/2\nonumber\\
\mbox{t=3\hspace{1ex}}
	& \mbox{apply $\mathbf{H}$}
	& \longrightarrow(\ket{{-2},0} + \ket{{-2},1}
			       + \ket{0,0} - \ket{0,1}\nonumber\\
 &	&       \null\hspace{9.3em} + \ket{0,0} + \ket{0,1}
			+ \ket{2,0} - \ket{2,1})/\sqrt{8}\nonumber\\
 &	& \mbox{\null\hspace{0.7em}} = (\ket{{-2},0} + \ket{{-2},1}
			       + 2\ket{0,0}
			       + \ket{2,0} - \ket{2,1})/\sqrt{8}\nonumber\\
 & \mbox{apply $\mathbf{S}$}
	& \longrightarrow(\ket{{-3},0} + \ket{{-1},1}
			       + 2\ket{{-1},0}
			       + \ket{1,0} - \ket{3,1})/\sqrt{8}
\label{eq:3steps}
\end{eqnarray}
In step 3, the component with the coin in state $\ket{1}$ at the origin
is eliminated while the component with coin state $\ket{0}$ is enhanced.
This is such a simple effect, entirely due to the wave nature of
quantum mechanics, that the need for a quantum system to gain the speed up,
rather than a wave system such as classical light,
has been questioned by \cite{knight03a}.  Indeed, the experiment with
classical light has
already been done by \cite{bouwmeester99a}.

To properly address this question, we need to distinguish two
different contexts: physical systems, and algorithms.
Table \ref{tab:algovsphys} shows examples of random walks in various
settings.
\renewcommand{\arraystretch}{1.5}
\begin{table}
\caption{Examples of random walks.}
\begin{center}
\label{tab:algovsphys}
\begin{tabular}{l|l|l|}
 &
        \null\hspace{2em}{\textsc{quantum}} 
&
        \null\hspace{2em}{\textsc{classical}}
\\
\hline
 {\textsc{physical}} &
         {atom in optical lattice} &
         {snakes and ladders \textit{(board game)}} \\
\hline
 {\textsc{computer}} &
         {glued trees algorithm} &
         {lattice QCD calculation}\\
\hline
\end{tabular}
\end{center}
\end{table}
Lined up side by side like this, the distinctions between them
should be obvious, but notice that classical computer simulations
of all four examples have also been done.
\cite{dur02a} used simulation to assess the 
effects of errors when making an atom in an optical lattice perform
a quantum walk.  A quick search on the Internet finds a number of 
online snakes and ladders games you can play
(e.g., 
\texttt{\small http://www.bonus.com/bonus/card/SnakesandLadders.html}).
Simulations of quantum walk algorithms are described in \cite{tregenna03a},
and the one that seems silliest
-- classical simulation of a classical computer algorithm --
is actually the most useful, for development of 
software to run on new parallel computers for lattice QCD
calculations, \cite{boyle02a}.

For physical systems we can give a straightforward answer to the
question of what makes a quantum walk `quantum': like all
quantum systems they should exhibit complementarity
\citep{bohr28b,bohr50a}.  The text-book complementarity experiment
is Young's double-slits, in which quantum particles (such as photons
or electrons) are thrown through a pair of closely spaced slits, 
with a device that can detect which slit they pass through.
If the particles pass through the slits unobserved, an interference
pattern builds up from many particles arriving
sequentially, conversely, if we observe which slit they pass through, the
interference pattern disappears, corresponding to the wave and particle
natures of quantum particles respectively.
Although early descriptions of complementarity concern mutually
incompatible measurements, \cite{wootters79a} presented a
description
in which complementarity can be quantified
as a trade-off between knowledge about which way each particle
goes vs the sharpness of the interference pattern.

\begin{figure}
    \begin{center}
	\resizebox{0.6\textwidth}{!}{\includegraphics{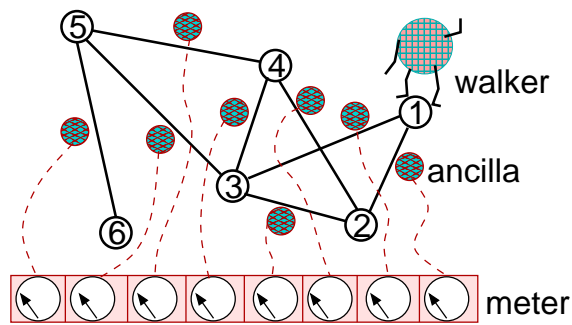}}
    \end{center}
    \caption{Quantum walker on a general graph with ancillae to
	measure the path.}
    \label{fig:ancillagraph}
\end{figure}

Quantum walks can be viewed as just a more complicated set of paths than
a double-slit, so, to demonstrate complementarity in
a quantum walk, we need to show the trade-off between quantum
interference and knowledge of the path taken by the walker.
This is just an extension of what was used in section \ref{sec:decoher}
to show a quantum walk reduces to a classical random walk when decohered.
We modeled decoherence by
applying measurements of the path with probability $p$.
If we now view $p$ as a coupling strength between our measuring device and
the walker, we are doing a weak, or partial, measurement through which
we learn only partial information about the path of the walker.
\cite{kendon04a} discuss in detail how to do this, using ancillae to
couple the measuring device to the walker, as shown in
figure~\ref{fig:ancillagraph}.

An optical system can also exhibit complementarity,
since light is not classical, it is made up of many photons.
Photons do not interact with each other, so we can consider each photon
in a `classical' light beam as doing an independent quantum walk.
The optical quincunx experiment \citep{bouwmeester99a}
is thus a quantum walk experiment in which only the wave nature of light
is demonstrated.  To also demonstrate the particle nature of light
would not be easy, but a possible approach using very low light
levels (essentially one photon at a time) is discussed in \cite{kendon04a}.
On the other hand, water waves could only reproduce the wave nature
of a quantum walk, since the underlying particles are not behaving
independently, and the wave nature of water arises out of their collective
behaviour.

\section{Algorithmic efficiency}
\label{sec:simeff}

We will now consider the question of what makes a quantum walk algorithm
more efficient than a classical algorithm.
A detailed answer is beyond the scope of this paper, being
the substance of a major branch of computer science,
but we can give a flavour of what is involved.
An algorithm is an abstract mathematical concept, 
whereas we want to consider how efficiently we can run an algorithm on a
computer of some sort.  Computer scientists do this by associating
a cost with each step of the algorithm and with the amount of memory
required, embodying the idea that
physical computers have a finite size (memory) and work at a finite
rate of elementary calculation steps per unit time.
This gives us a way to determine if one algorithm is intrinsically
faster than another.
However, if we want to compare algorithms run on quantum or classical
computers, we also need to compare computers:
the fundamental way to do this is to ask
whether one computer can simulate another with only a constant extra
overhead in computation.  If so, then they are equivalent for
computer science purposes (though not necessarily for practical
purposes).
In most cases, `efficient' means the algorithms or computers are
equivalent to within a factor that scales as a simple polynomial
(like $N^2$) of the size $N$ of the system, or the number of steps
it takes to run the algorithm.

If we could simulate a quantum 
algorithm efficiently on a classical computer, we would
immediately have a classical algorithm that is sufficiently
good to render the quantum algorithm superfluous.
This is a criterion regularly applied in quantum computing,
e.g. the Clifford group of operations can all be 
efficiently classically simulated, \cite{aaronson04a},
so they are not sufficient to build a quantum computer.
Quantum algorithms with no efficient classical simulation may
still be equaled or bettered by a classical algorithm using a
different method.
To prove an algorithmic speed up for a quantum algorithm,
we have to prove that no classical algorithm of any type can do as well:
such proofs are in general very difficult.  \cite{childs02a} do
prove their quantum algorithm is faster than any classical algorithm,
but Shor's algorithm \citep{shor95a} is only the `best known' algorithm
for factoring, there is no proof something faster can't be found in future.

\subsection{Simulation of classical systems}
\label{sec:csim}

Broadly speaking, a computer simulation of a physical system
is useful if it can be calculated significantly faster and with less
resources than just observing the physical system itself.  For
example, calculating the trajectory of a space probe
that takes five years to reach its destination is crucial: we cannot
just launch many test probes and see where they end up in order to
work out which trajectory we want.
Conversely, simulation of fluid flow is not 
so efficient, and wind tunnel testing is often used for aircraft
and turbine design.

One key difference between a classical
physical system and a computer simulation of it is that we represent
it by binary numbers in a register in the computer.
Thus, for our random walks on a line,
we need only $\log_2 N$ bits in our register for
each $N$ points on the line.
This is an exponential saving in resources.
We still need to do $T$ steps of our random walk, so there is no saving
in the running time of the program, but we can find out where the
walker ended up by making just $log_2 N$ measurements, each with a
binary outcome (zero or one), in contrast to the physical system where
we potentially have to examine all of the $N$ positions on the line
to discover where the walker ended up.
This binary encoding in a computer, compared to unary in a physical
system, applies equally to classical and quantum digital computers.
In the quantum context it was first remarked on by \cite{Jozsa98a},
and \cite{blume02a} presented explicit calculations of the most
efficient way to use quantum systems as registers
(they don't need to be binary).

Of course, we don't get something for nothing: there is a trade off for
the exponentially smaller resources of the binary encoding.
When we update the state
of our random walker, we know that it can only move one step at a time.
On the physical line, these steps are to nearest neighbouring points,
but in the computer register, going from position `7' to position `8'
is $0111\longrightarrow 1000$ in binary, so every one of the bits has
to flip, the elementary operations are no longer `local' to a small
portion of the space.  The price we pay (flipping up to $log_2N$ bits compared
to a single hop on the line) is still in general a better deal than using
exponentially more space to accommodate the whole $N$ points.

\subsection{Simulation of quantum systems}
\label{sec:qsim}

The reason why quantum systems are generally hard to simulate with
a classical computer is because of the quantum superpositions that
all have to be kept track of, the original
motivation for quantum computing.
The Hilbert space of a quantum system is exponentially bigger than the 
phase space of a classical system with the same number of
degrees of freedom.  The numbers get fantastically large.  A ten
(classical) bit register can be in any one of $2^{10}$ different states.
A ten qubit register can be in superposition of all of those
$2^{10}$ states at once, with any proportion you like of each state.
The quantum computer with a ten qubit register can thus perform up
to $2^{10}$ calculations in parallel, each corresponding to a different
input state.
A hundred qubit computer can do
$2^{100}$ calculations at the same time, more than there are particles in
the universe (estimated to be around $2^{87}$).
Again, of course, there is a trade off for this enormous parallelism.
You only get \emph{one} answer out, not $2^{10}$.  A classical computer
simulation of a ten qubit quantum computer is easily done with today's
computing power, and it can tell you what the whole superposition
is at any stage of the computation, including all $2^{10}$ possible
answers, so it is significantly more
powerful in terms of the information it makes available.

A quantum simulation of a quantum system can be done efficiently.
This is not a trivial statement.  It is necessary to show, as
\cite{lloyd96a} did, that if one maps the Hilbert space of the
quantum system directly onto the Hilbert space of the quantum
register (no binary encoding needed here), then it is possible to
simulate the Hamiltonian evolution of the quantum system to
sufficient accuracy using a sequence of standard Hamiltonians
applied to the quantum register, each for short period of time.
Quantum simulation has already been demonstrated \citep{somaroo99a},
for small systems using NMR quantum computers.  However, like
classical analogue computing, quantum simulation has a problem
with accuracy, which does not scale efficiently with the
time needed to run the simulation, \cite{brown06a}.

\subsection{Analogue computing}
\label{sec:analogue}

The discussion up until now has implicitly assumed we are discussing
digital computers, but this leaves out an important piece of the puzzle.
Classical analogue computers go back to \cite{shannon41a}, who showed
how they can solve differential equations, given the boundary conditions
as inputs.
Using a system of non-linear boxes with lots of feedback,
they can do this very efficiently; general purpose analogue computers
can be constructed with a small number of such boxes that will 
approximate any differential equation to arbitrary accuracy,
\cite{rubel81a}.  It is also possible to extend
the original concept \citep[e.g.,][]{rubel93a}
to solve a wider class of equations.
Analogue computers do not binary encode their data, e.g., the inputs
might be directly related to the size of a voltage, 
and the answer is obtained by measuring the outputs of a suitable
subset of the non-linear boxes.  The answer is thus as accurate as
it is possible to make these measurements, and extra accuracy has
an exponential cost compared to digital computing.
\cite{rubel89a} proved that a classical digital
computer can simulate an analogue computer, but the reverse question of
whether an analogue computer can simulate a digital computer efficiently
is open.
The utility of analogue computers is in their speed: though they
are now rare given the power of today's digital computers,
research continues, and the combination of an analogue `chip' in a
digital computer to perform specific real-time operations such as
video rendering is one of many possible new applications.

The relevance of analogue computing to quantum computing has not been
fully explored, \citep{jozsa05b}.
Quantum systems accomplish their feats of superposition
through an analogue quality: a qubit may be in any superposition of
zero and one,
\begin{equation}
\ket{\psi} = \alpha\ket{0} + \beta e^{i\phi}\ket{1},
\end{equation}
where $\alpha$ and $\beta$ are real numbers satisfying $\alpha^2 + 
\beta^2 = 1$ and $0\le\phi<2\pi$.
The quantum operations we apply to perform a quantum computation alter
the values of $\alpha$, $\beta$ and $\phi$, and we can only do this to
a finite accuracy.  For digital quantum computing, analysis of the
effects of limited accuracy in the quantum gates 
\citep[for example, ][on Shor's algorithm]{nielsenchuang00} 
suggest it does not affect the computation significantly, but 
as already mentioned, for quantum simulation the errors scale less favorably
(linear in the size of the problem, \cite{brown06a}).

\section{The future of quantum walk algorithms}
\label{sec:outlook}

We do not yet know if we can build a quantum computer large enough to
solve problems beyond the reach of the classical computational power
available to us, but the rewards for success are so exciting that the
challenge is well worth the sustained effort of many years of research
by hundreds of talented scientists around the world.  And should we find
out it is fundamentally impossible to build such a machine, that in
itself will tell us crucial facts about the way our universe works.

The first useful application for quantum computers is likely to be
simulation of quantum systems.  Small quantum simulations have
already been demonstrated, and useful quantum simulations -- that
solve problems inaccessible to classical computation -- require
less qubits than useful factoring, and may be achievable within
five to ten years.
A useful digital quantum computer for factoring
will need to create and maintain a complex
superposition of at least a few thousand qubits. 
\cite{aaronson03a} provides a quantitative estimate of what this might mean, 
against which we can measure experimental achievements.
Such a computer is probably ten to twenty years away from realization.

However, quantum walk algorithms may be among the
first algorithms to provide useful computation beyond quantum
simulation, both because of the
wide range of important problems they can be applied to, and their
versatility for optimizing their performance.
In making this prediction I have a parallel to draw on from 
classical algorithms, in the development of lattice gas methods
for simulating complex fluids.  The original idea was first 
published by \cite{hardy73a}, but had a major shortcoming: on a square or 
cubic lattice, too many quantities are conserved.
The solution (add diagonals to the lattice)
came fifteen years later from \cite{frisch87a},
when the development of computers
meant that testing and actual use of such methods was within
reach.  Practical computational methods
proceed with a mixture of theory and refinement through testing
real instances on real computers.  For quantum walks we only have the
theory, and simulation on classical computers.  This is not enough
to fully develop useful applications, and I expect the availability of
a working quantum computer
to speed up the development of quantum walk algorithms significantly.

\begin{acknowledgements}
I thank many people for useful and stimulating discussions of quantum walks,
Andris Ambainis,
Dorit Aharonov,
Sougato Bose,
Ivens Carneiro,
Andrew Childs,
Richard Cleve,
Jochen Endrejat,
Ed Farhi,
Will Flanagan,
Mark Hillery,
Peter H\o{}yer,
Julia Kempe,
Peter Knight,
Barbara Kraus,
Meng Loo,
Rik Maile,
Olivier Maloyer,
Cris Moore,
Eugenio Rold{\'a}n,
Alex Russell,
Barry Sanders,
Mario Szegedy,
Tino Tamon,
Ben Tregenna,
John Watrous,
and
Xibai Xu.
VK is funded by a Royal Society University Research Fellowship.
\end{acknowledgements}

\small

\label{lastpage}
\end{document}